# CdS Nanocrystallines: Synthesis, Structure and Nonlinear Optical Properties

Oleksandr Yanchuk
Department of Inorganic and Physical Chemistry,
Lesya Ukrainka Eastern European National University
Lutsk, Ukraine
yanchuk59@gmail.com

Oleg V. Marchuk
Department of Inorganic and Physical Chemistry,
Lesya Ukrainka Eastern European National University
Lutsk, Ukraine
Oleg_M_1974@i.ua

Galyna Myronchuk
Department of Inorganic and Physical Chemistry,
Lesya Ukrainka Eastern European National University
Lutsk, Ukraine

Iryna A. Moroz
Department of Materials Science,
Lutsk National Technical University
Lutsk, Ukraine
moroz.iryna1@gmail.com

Nazariy Andrushchak
Department of Computer-Aided Design Systems
Lviv Polytechnic National University
Lviv, Ukraine
nazariy.a.andrushchak@lpnu.ua

Oleksii A.Vyshnevskyi
M.P.Semenenko Institute of Geochemistry, Mineralogy and Ore Formation
NAS of Ukraine
Kyiv, Ukraine
vyshnevskyy@i.ua

Iwan V. Kityk
Faculty of Electrical Engineering,
Czestochowa University of Technology
Czestochowa, Poland

Andriy V. Kityk
Faculty of Electrical Engineering,
Czestochowa University of Technology
Czestochowa, Poland
andriy.kityk@univie.ac.at

Katarzyna Ozga
Faculty of Electrical Engineering,
Czestochowa University of Technology
Czestochowa, Poland
cate.ozga@wp.pl

Jaroslaw Jedryka
Faculty of Electrical Engineering,
Czestochowa University of Technology
Czestochowa, Poland
jaroslaw.jedryka@o2.pl

Artur Wojciechowski
Faculty of Electrical Engineering,
Czestochowa University of Technology
Czestochowa, Poland

Anatoliy Andrushchak
Department of Telecommunication
Lviv Polytechnic National University
Lviv, Ukraine
anatolii.s.andrushchak@lpnu.ua

*Abstract* — We report the synthesis, structure and nonlinear optical properties of cadmium sulphide (CdS) nanocrystallines (NCs) synthesized electrochemically both with and without detergent ATLAS G3300. Relevant structural and morphological features are explored by X-ray diffraction and scanning electron microscopy (SEM) techniques. The efficiency of the second harmonic generation (SHG) appears to be strongly dependent on the energy density of the incident fundamental laser radiation and NC sizes.

## I. Introduction

Semiconductor nanocrystals have diverse practical applications, including optoelectronic transistor components and fluorescent biological labels [1-4]. Relevant materials exhibit good photomechanical features for laser operation basically due to good photo-thermal stability against external laser radiation [5] combined with high phonon anharmonicity being typical for chalcogenides [6-7]. However, there remains still a problem in obtaining low size dispersed samples. The optical and electronic features of nanocrystals are dependent on materials-scale properties, such as particle size distribution and shape [8]. Therefore, the challenge in semiconductor nanocrystal synthesis lies in strict control of relevant characteristics by manipulating synthetic parameters. Among existing methods, the electrochemical electrolysis appears to be very promising representing rather flexible and relatively simple technique [9]. Besides, different techniques for investigation of properties of new materials in optical [10,11] and millimeter-wave range [12-14] as well as their application for electro-optics and photonics [15,16] can be applied.

In this work, we compare the morphology of cadmium sulfide (CdS) nanocrystallines (NCs) synthesized electrochemically both with and without ATLAS G3300 detergent, i.e. employing so-called detergent or detergentless electrolytic synthesis methods, respectively.

Section 2 describes the methodology of CdS NC synthesis, experimental techniques applied in structural studies of CdS NCs, such as X-ray powder diffraction and scanning electron microscopy, as well as the nonlinear optical setup used in investigations of the laser stimulated second harmonic generation (SHG) induced by NCs. Section 3 presents the structural and morphological features of CdS NCs as well as their second-order nonlinear optical properties explored in SHG experiments.

## II. Experimental

CdS NCs were produced by the electrolysis of aqueous sodium chloride (58.442 g/L or 1.0 mol/L) and thiourea (15.22 g/L or 0.20 M) solution with a soluble cadmium anode at ($T$ = 363.2 K) at different current densities. In our experiments, the electrolysis lasted for 20 minutes. The principal parameter in such synthesis is the current density specified in Tables 1 and 2. ATLAS G3300 *(Isopropylamine dodecyl benzenesulfonate)* was used to control the aggregate stability of the aqueous suspensions. In this case, sodium chloride, thiourea and ATLAS G3300 were placed into a volumetric flask (1 liter) and dissolved in hot distilled water. The contents of the flask were brought to the constant temperature of 90 °C stabilized by a thermostat. The obtained solution was transferred to a 400 ml electrolyzer, and cylindrical cadmium electrodes with a surface area of about 5 cm$^2$ were it was immersed into the solution and connected to a power source (B5-46 AC/DC converter).

TABLE I. Current density in the detergentless electrochemical synthesis of the cadmium sulphide NCs

| Sample number | B01 | B03 | B05 | B07 |
|---|---|---|---|---|
| Current density $j$, A/cm$^2$ | 0.23 | 0.32 | 0.40 | 0.48 |

TABLE II. Current density in the detergent (ATLAS G3300, 1.0 g/L) electrochemical synthesis of the cadmium sulphide NCs

| Sample number | D06 | D08 | D10 | D12 |
|---|---|---|---|---|
| Current density $j$, A/cm$^2$ | 0.23 | 0.32 | 0.40 | 0.48 |

Each experiment required about 200 ml of the prepared solution. The electrolyzer with the immersed electrodes and a thermometer were placed into a thermostat. After heating to the desired temperature, the timer was switched on and the power source was set to specific current and voltage values. Upon anode dissolution, orange particles immediately began to grow around it. When the synthesis of CdS NCs reached saturation, the electrolyzer was removed from the thermostat and left to cool down to ambient temperature.

After cooling, the contents of the glass, in which the electrolysis took place, was poured into a larger glass with distilled water and left until the next day to completely settle the powder. The next day the powder settled on the bottom of the glass and the remaining solution have been poured into a glass with distilled water. The resulting suspension was left to defend until the next day. The powdered precipitate has been washed repeatedly until the solution, which was poured from the glass, stopped foaming. The remaining precipitate was then transferred to a Petri dish and dried at 50 °C. The cadmium electrodes were washed in the distilled water and then dried in air and weighed [9].

Synthesized orange powders were characterized by scanning electron microscopy and X-ray powder diffraction (diffractometer DRON-4-13, Cu-Kα radiation). The morphology and size of synthetic CdS nanocrystals were determined by using a field emission scanning electron microscope JSM-6700F ("JEOL") at IGMOF NASU. Samples were coated by a platinum film of 30 Å thickness applying the sputtering method. SEM images were recorded in SE mode applying 10 kV accelerating voltage and a beam current of 0.65 nA.

The experimental set-up used in our nonlinear optical measurements is presented in Fig. 1. Such a laboratory system has been used to detect the second harmonics of the light (SHG) generated by measured NC samples. As a source of fundamental radiation, the Nd:YAG nanosecond pulsed laser (pulse duration: 8 ns, wavelength: 1064 nm, frequency repetition: 10 Hz) was used. The power of the incident fundamental laser radiation was tuned by Glan's polarizer with a laser damage power density of 4 GW/cm$^2$.

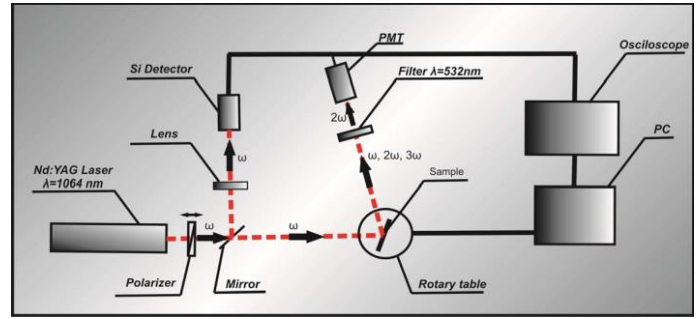

Fig. 1. The laboratory set-up for measurements of the second harmonic of the light.

The laser beam has been characterized by a profile diameter equals to about 8 mm and the maximal energy density of about 100 J/m$^2$. The energy of the incident (fundamental) laser radiation was measured by the germanium photodetector whereas the SHG light has been detected by a Hamamatsu photomultiplier supplied by an interference filter with a maximum transmission at 532 nm and spectral width of about 5 nm providing a selection of the frequency-doubled electromagnetic radiation generated by the NC samples. Measured NC powders were fixed on microscope glass slides which then have been placed on a rotating table controlled by PC.

The intensities of detected fundamental and second harmonic signals were monitored and analyzed using a Tektronix MSO 3054 oscilloscope (sampling: 2.5 GS) connected with PC via a USB interface for relevant data acquisition. The entire measuring stand was placed under the box eliminating the influence of external undesirable light scattering.

## III. Results and Discussion

XRD diffractograms for all eight CdS NC samples are presented in Figs. 2, 3. Reference diffraction powder patterns corresponding to *F43m* and *P6$_3$mc* CdS crystal structures are indicated by blue and orange lines, respectively, for comparison with recorded XRD patterns of CdS NCs, see labelled. The relevant analysis shows that synthesized CdS NCs are characterized by either hexagonal or cubic structures corresponding to wurtzite or sphalerite type, respectively. An approximate content of wurtzite and sphalerite NCs is given in Table. III.

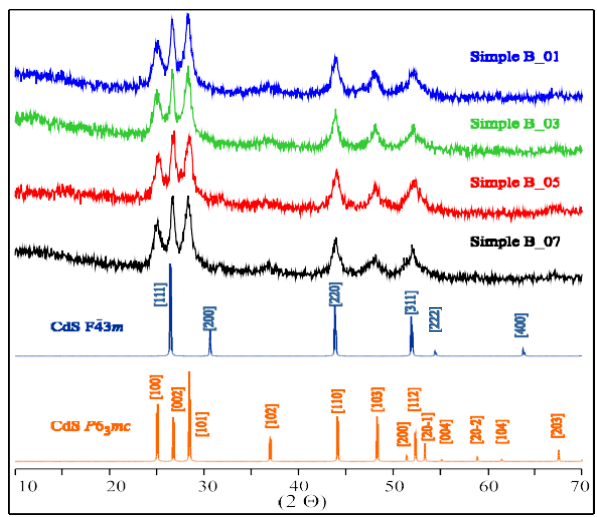

Fig. 2. XRD patterns of sediment powder of CdS NCs obtained by electrolysis of sodium chloride and thiourea with cadmium anode at a temperature of 90 °C, electrolysis time 20 min and different current densities (A/cm$^2$): sample B01 – 0,23, sample B03 – 0.32, sample B05 – 0.40, sample B07 – 0.48. Reference XRD powder patterns of $F\bar{4}3m$ and $P6_3mc$ CdS crystal structures are presented for comparison, see labeled.

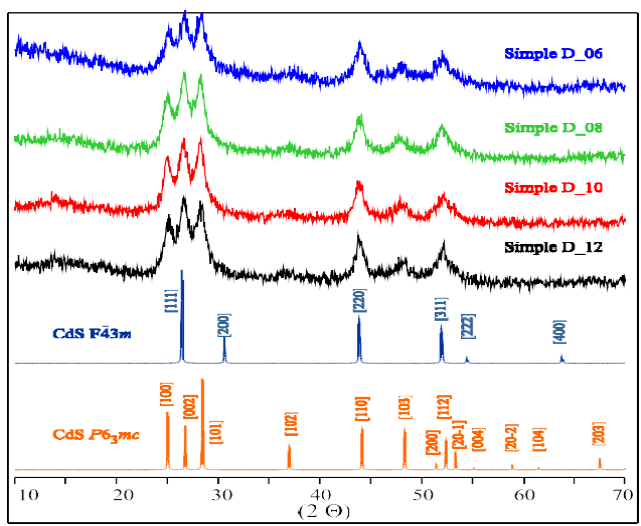

Fig. 3. XRD patterns of sediment powder of CdS NCs obtained by electrolysis of sodium chloride and thiourea with cadmium anode at a temperature of 90 °C, electrolysis time 20 min in the presence 1 g/L ATLAS G3300 and different current density (A/cm$^2$): sample D06 – 0,23, sample D08 – 0.32, sample D10 – 0.40, sample D12 – 0.48. Reference XRD powder patterns of $F\bar{4}3m$ and $P6_3mc$ CdS crystal structures are presented for comparison, see labeled.

TABLE III. THE CALCULATED APPROXIMATE CONTENT OF WURTZITE AND SPHALERITE NCs IN CADMIUM SULPHIDE POWDER

| № samples | $P6_3mc$, % | $F\bar{4}3m$, % | Current density $j$, A/cm$^2$ |
|---|---|---|---|
| B01 | 77 | 23 | 0.23 |
| B03 | 84 | 16 | 0.32 |
| B05 | 80 | 20 | 0.40 |
| B07 | 75 | 25 | 0.48 |
| D06 | 74 | 26 | 0.23 |
| D08 | 73 | 27 | 0.32 |
| D10 | 75 | 25 | 0.40 |
| D12 | 73 | 27 | 0.48 |

CdS of wurtzite structure ($P6_3mc$) evidently dominates here. Moreover, an applying of ATLAS G3300 detergent in NC synthesis results in the practically constant ratio (3:1) of wurtzite to sphalerite NCs. The current density evidently influences the ratio of wurtzite and sphalerite modifications in synthesized CdS powders, mainly in the case of detergentless synthesis. The electrolysis current density, on the other hand, practically does not affect the ratio of wurtzite and sphalerite modifications in the resulting CdS powders whenever the detergent ATLAS G3300 is applied in relevant electrolysis process.

Figs. 4 and 5 demonstrate SEM images of CdS NC powders. One may see that the synthesized CdS NCs are predominantly of flakes form.

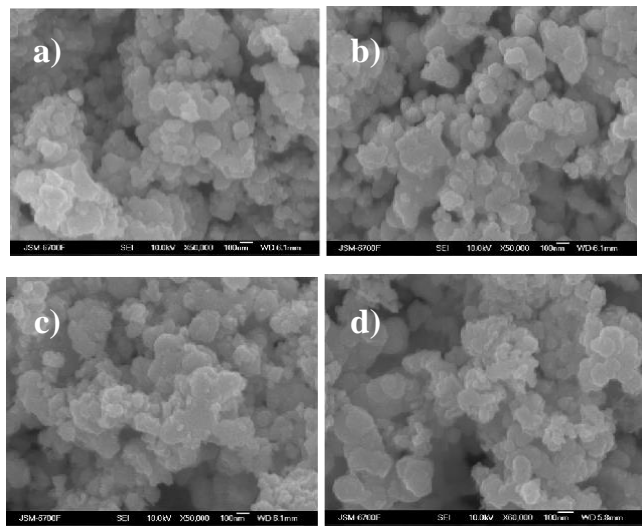

Fig. 4. SEM images samples of cadmium sulfide synthesized under the same temperature (90 ºC), electrolysis time (20 min), concentrations of NaCl (1.0M) and thiourea (0.2M), but different electrolysis current density (A/cm$^2$): sample B01 – 0.23, sample B03 – 0.32, sample B05 – 0.40 and B07 – 0.48.

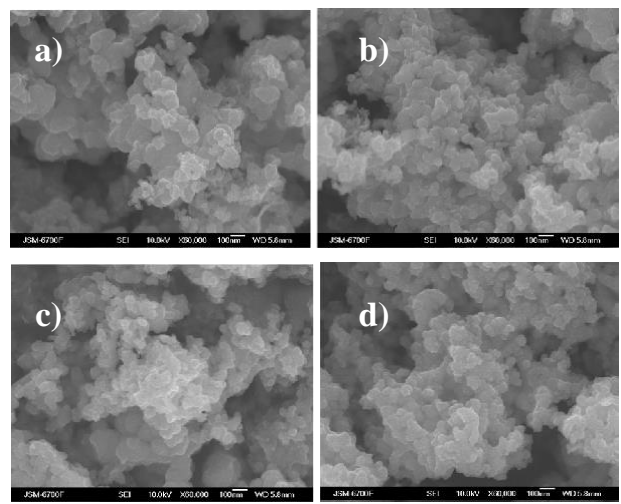

Fig. 5. SEM images samples of cadmium sulfide synthesized under the same ATLAS content (1.0 g/L), temperature (90 ºC), electrolysis time (20 min), concentrations of NaCl (1.0M) and thiourea (0.2M) but different electrolysis current density (A/cm$^2$): a) sample D06 – 0.23, b) sample D08 – 0.32, c) sample D10 – 0.40 and d) D12 – 0.48.

The powder particles obtained within the detergentless electrolysis synthesis (B01, B03, B05, B07) have sizes in a quite wide range, from 5 to 70 nm (Fig 4).

The diameter of the flake's ranges from 15 to 70 nm and the thickness from 5 to 25 nm. The average particle diameters appear in the range of 35-40 nm. The synthesis current density, which ranges from 0.232 to 0.48 A/cm2, has negligible influence on the particles-flakes diameters. The powder particles obtained within the detergent electrolysis synthesis (D06, D08, D10, D12) are of size from 5 to 50 nm (Fig.5). The diameter of the flakes appears in the range from 15 to 50 nm and the thickness of 5÷15 nm. The influence of current density (0.23÷0.48 A/cm$^2$) on the NCs sizes is also negligible. Taking together, in the cases of both detergent and detergentless elecrolysis synthesis one deals with nanoscale CdS particles.

The average particle diameters appear in the range of 24.0-26.6 nm. The distribution of the number of particles (as a percentage of the total number of particles) in the four-stage scale (range 10÷50 nm) indicates that the largest number of particles appear in the range from 21 to 30 nm (Fig. 6).

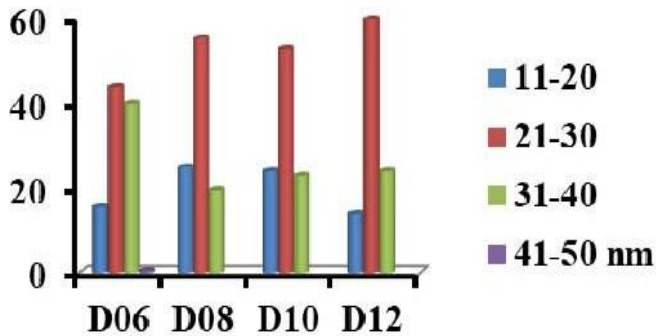

Fig. 6. The distribution of CdS NCs over sizes (as a percentage of the total number of particles) in the four-stage scale (range 10-50 nm).

The spread of particles in size, as seen from Fig. 5, is insignificant. This is facilitated by the detergent addition to the electrolyte. Accordingly, one may conclude that ATLAS 3300 represents here as good particle size stabilizer.

Among the laser operated optoelectronic features we are concentrated on the studies of the laser-induced second harmonic generation (SHG) representing the second-order nonlinear optical effect described by the third rank polar tensor. The SHG efficiencies versus the fundamental laser energy density are presented in Figs. 7 and 8 for CdS NCs synthesized electrochemically at different current densities within both detergentless and detergent (ATLAS G3300, 1.0 g/L) electrolysis synthesis, respectively. One may see that the grain size exhibits rather a significant effect on the intensity of the SHG signal. Fractions of finer powder particles give lower SHG signals.

I. CONCLUSIONS

Novel CdS NCs were synthesized by detergentless and detergent (ATLAS 3300) electrolysis methods. XRD analysis suggests that the CdS powder content is dominated by NCs of wurtzite type. CdS powders obtained by the detergent electrolysis method are characterized by practically constant ratio (3:1) of wurtzite to sphalerite NCs. The synthesized CdS NCs are predominantly of flakes form.

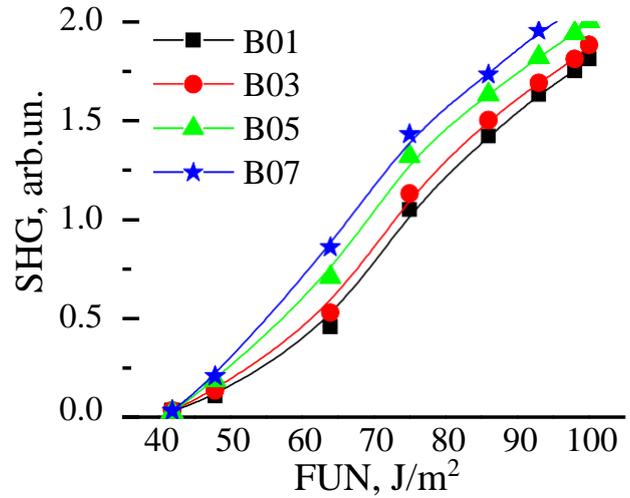

Fig. 7. Dependence of the SHG efficiency versus the fundamental laser energy density of the CdS NC powders synthesized at different current densities by detergentless electrolysis method.

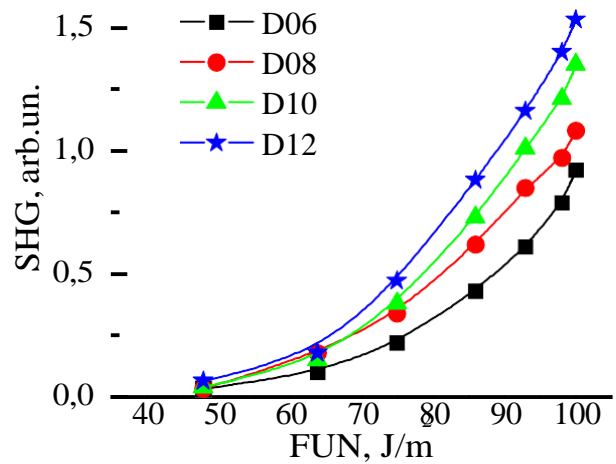

Fig. 8. Dependence of the SHG efficiencies versus the fundamental laser energy density of the CdS NC powders synthesized at different current densities by detergent (ATLAS G3300, 1.0 g/L) electrolysis method.

The influence of current density on the NCs size is negligible. The average particle diameters appear in the range of 24.0-26.6 nm, whereas the sizes of most particles are in the range from 21 to 30 nm. The detergent ATLAS 3300 represents a good particle size stabilizer. The average size of CdS NCs synthesized by the detergent electrolysis method statistically is smaller than the average particle size obtained by counterpart detergentless method. Laser-induced SHG has been studied for CdS NCs obtained by both electrolysis methods. The SHG light intensity is found to be significantly dependent on the NC size.

ACKNOWLEDGMENT

The presented results are part of a project that has received funding from the European Union's Horizon 2020 research and innovation programme under the Marie Skłodowska-Curie grant agreement No 778156. Support from resources for science in years 2018-2022 granted for the realization of international co-financed project Nr W13/H2020/2018 (Dec. MNiSW 3871/H2020/2018/2) is also acknowledged.

REFERENCES

[1] J.K. Jaiswal, H. Mattoussi, J.M. Mauro and S.M. Simon, "Long-term multiple color imaging of live cells using quantum dot bioconjugates," Nat. Biotechnol, vol. 21, pp. 47-51, December 2003.

[2] K.R. Brown, D.A. Lidar and K.B. Whaley, "Quantum computing with quantum dots on quantum linear supports," Phys. Rev., vol. A65, 012307, January 2002.

[3] W.C.W. Chan, D.J. Maxwell, X.H. Gao, R.E. Bailey, M.Y. Han and S.M. Nie, "Luminescent quantum dots for multiplexed biological detection and imaging," Curr. Opin. Biotechnol, vol. 13, pp.40-46, February 2002.

[4] R.Y. Sweeney, C. Mao, X. Gao, J.L. Burt, A.M. Belcher, G. Georgiou and B.L. Iverso, "Bacterial biosynthesis of cadmium sulfide nanocrystals," Chem. Biol., vol. 11, pp.1553-1559, November 2004.

[5] K. Ozga, O.M. Yanchuk, L.V. Tsurkova, O.V. Marchuk, I.V. Urubkov, Y.E. Romanyuk, O. Fedorchuk, G. Lakshminarayana and I.V. Kityk, "Operation by optoelectronic features of cadmium sulphide nanocrystallites embedded into the photopolymer polyvinyl alcohol matrices," Appl. Surf. Sci., vol. 446, pp. 209-214, July 2018.

[6] O.V. Parasyuk, V.S. Babizhetskyy, O.Y. Khyzhun, V.O. Levytskyy, I.V. Kityk, G.L. Myronchuk, O.V. Tsisar, L.V. Piskach, J. Jedryka, A. Maciag and M. Piasecki, "Novel quaternary $TlGaSn_2Se_6$ single crystal as promising material for laser operated infrared nonlinear optical modulators," Crystals, vol.7, pp. 341-357, November 2017.

[7] A.S. Krymus, G.L. Myronchuk, O.V. Parasyuk, G. Lakshminarayana, A.O. Fedorchuk, A. El-Naggar, A. Albassam and I.V. Kityk, "Photoconductivity and nonlinear optical features of novel $Ag_xGa_xGe_{1-x}Se_2$ crystals," Mater. Res. Bull. vol. 85, pp. 74-79, January 2017.

[8] . Ebothé, J. Michel, I.V. Kityk, G. Lakshminarayana, O.M. Yanchuk and O.V. Marchuk, "Influence of CdS nanoparticles grain morphology on laser-induced absorption, " Physica E, vol 100, pp. 69-72, June 2018.

[9] Patent 93471U Ukraine, IPC (2014.01) C 01G 11/00, Method for obtaining nanopowders of cadmium sulfide by electrolysis of an aqueous solution of indifferent electrolyte, Applicant and Patent Owner of Lesya Ukrainka Eastern European National University. – No. u201313037; Stated. 11.11.2013; Has published 10.10.2014, Byul. No. 9. - 3 p.

[10] Y. Shchur, A. Andrushchak, V. Strelchuk, A. Nikolenko, V. Adamiv, N. Andrushchak, P. Göring, P. Huber, "$KH_2PO_4$ + Host Matrix (Alumina/$SiO_2$) Nanocomposite: Raman Scattering Insight," 2019 21st International Conference on Transparent Optical Networks (ICTON), Angers, France, 2019, pp. 1-4.

[11] A.S. Andrushchak, T.I. Voronyak, O.V. Yurkevych, N.A. Andrushchak, A.V. Kityk, "Interferometric technique for controlling wedge angle and surface flatness of optical slabs," Optics & Lasers in Engineering, vol. 51, No.4, pp.342-347, 2013.

[12] N. Andrushchak, I. Karbovnyk, "LabVIEW-Based Automated Setup for Interferometric Refractive Index Probing," SLAS Technology: Translating Life Sciences Innovation, pp.1-7, 2019.

[13] N.A. Andrushchak, O.I. Syrotynsky, I.D. Karbovnyk, Ya.V. Bobitskii, A.S. Andrushchak, A.V. Kityk, "Interferometry technique for refractive index measurements at subcentimeter wavelengths," Microwave and Opt. Technol. Lett, vol. 53, pp. 1193–1196, 2011.

[14] N.A. Andrushchak, I.D. Karbovnyk, K. Godziszewski, Ye. Yashchyshyn, M.V. Lobur, A.S. Andrushchak, "New Interference Technique for Determination of Low Loss Materials Permittivity in the Extremely High Frequency Range," IEEE Trans. Instrum. Meas., vol. 64 (11), pp. 3005-3012, 2014.

[15] N. Andrushchak, P. Goering, A. Andrushchak, "Nanoengineering of Anisotropic Materials for Creating the Active Optical Cells with Increased Energy Efficiency," 2018 14th International Conference on Advanced Trends in Radioelectronics, Telecommunications and Computer Engineering, 2018, February 20-24, pp. 484-487.

[16] N. Andrushchak, B. Kulyk, P. Goering, A. Andrushchak, B. Sahraoui "Study of second harmonic generation in KDP/$Al_2O_3$ crystalline nanocomposite," Acta Phys. Pol. A, vol. 133, pp. 856-859, 2018.